\documentclass[]{aa}
\usepackage{graphicx}

\begin{document}
\title{Clues to the formation of the Milky Way's thick disk}


\author{M. Haywood\inst{1} \and  P. Di Matteo\inst{1} \and O. Snaith\inst{1,2} \and M.~D. Lehnert\inst{3}}

\authorrunning{Haywood et al.}

\offprints{M. Haywood, \email{Misha.Haywood@obspm.fr}}

\institute{GEPI, Observatoire de Paris, CNRS, Universit$\rm \acute{e}$  Paris Diderot, 5 place Jules Janssen, 92190 Meudon, France
\email{Misha.Haywood@obspm.fr}
\and 
Current Address: Department of Physics \& Astronomy, University of Alabama, Tuscaloosa, Albama, USA
\and
Institut d'Astrophysique de Paris, CNRS UMR 7095, Universit$\rm \acute{e}$
Pierre et Marie Curie, 98 bis Bd Arago, 75014 Paris, France}

\date{Received / Accepted}

\abstract{We analyse the chemical properties of a set of solar
vicinity stars, and show that the small dispersion in abundances of
$\alpha$-elements at all ages provides evidence that the SFH has been
uniform throughout the thick disk.  In the context of long time scale
infall models, we suggest that this result points either to a limited
dependence of the gas accretion on the Galactic radius in the inner
disk (R$<$10~kpc), or to a decoupling of the accretion history and
star formation history due to other processes governing the ISM in the
early disk, suggesting that infall cannot be a determining parameter of
the chemical evolution at these epochs.  We argue however that these
results and other recent observational constraints -- namely the lack
of radial metallicity gradient and the non-evolving
scale length of the thick disk -- are better explained if the early
disk is viewed as a pre-assembled gaseous system, with most of the gas
settled before significant star formation took place -- formally the
equivalent of a closed-box model.  In any case, these results point to
a weak, or non-existent inside-out formation history in the thick disk,
or in the first 3-5 Gyr of the formation of the Galaxy. We argue however
that the growing importance of an external disk whose chemical properties 
are distinct from those of the inner disk would give the
impression of an inside-out growth process when seen through snapshots
at different epochs. 
The progressive, continuous process usually invoked may not have actually existed in the Milky Way.}


\maketitle

\section{Introduction}

The cosmic star formation history shows that galaxies went through their
most vigorous phase of star formation at redshift z=1.5-3 \citep[e.g][]{mad14}.  
Structures thought to have formed at this epoch
in disk galaxies are classical bulges and thick disks.  The stellar
content of the nearby universe in environments with average density is
dominated by pure disk galaxies \citep[][]{kor10,fis10,lau14}.
This suggests that classical bulges represent only a minor mass component of galaxies.  Thick disks,
however, are found to be ubiquitous, and are detected on nearly all
nearby edge-on disks galaxies \citep[][]{com11}.  Recent results
\citep{nes13a,nes13b,pdm14a,pdm14b} have confirmed the view
that the Milky Way is also a pure disk galaxy \citep[][]{she10,kun12},  i.e., it has no substantial classical bulge. Further,
measurements of its star formation history have shown that during the
peak of SF in the universe, the MW was a star-bursting galaxy and was
forming its thick disk \citep[][]{sna14a,sna14b,leh14}.
The structural parameters of this population, with a short scale length
\citep[][]{ben11,che12a,bov12}
suggest it constitutes a major part of the mass in the
Milky Way's central region (Ness et al. 2013a,b, Di Matteo et al. 2014a,b,
Haywood et al. in preparation).

This emerging scheme suggests that the formation of the thick disk is
crucial phase in the formation of disks \citep[see][]{leh14},
both because this population could represent a substantial fraction of
the mass of galaxies and because it may account for the main increase in
the metal enrichment of galaxies since the formation of the first stars.
These views are opposite to the standard chemical evolution scenario based
on long term infall 
\citep[e.g][]{chi97,fra13}
and the recognition that the solar vicinity contains too few low metallicity
dwarf stars compared to the ``closed-box'' model. In other words,
observations of the solar vicinity would suggest that the enrichment
rate has been faster than the star formation rate.  Infall models have
been introduced in order to solve this problem by limiting the dilution
of ejected metals to a disk parsimoniously provisioned with gas, and by
boosting the increase of the metallicity at early times without creating
too many metal-poor stars.  Hence, these scenarios leave no room for
a substantial component of intermediate metallicity in the Milky Way.
This credo has been questioned by a new scheme presented in \cite{hay13}, \cite{sna14a, sna14b}
and \cite{hay14a, hay14b}.

Another key feature of the standard chemical evolution scenario is to make the accretion
of gas radially dependent, with slower accretion outwards, in order to mimic an inside-out formation. 
Inner regions form their stars first, because gas density increases faster, while outer
regions grow at later times and at a slower pace.  Because of the coupling
assumed between accretion and SFR in these models, a natural signature
of such scheme would therefore be that star formation histories vary
with distance from the galactic center. Direct measurement of the star
formation history outside the solar vicinity is not feasible currently,
however we know that a given SFH leaves specific signatures on the radial
and temporal dependence of the chemical abundances in stars \citep[][]{hay13, sna14a, sna14b}
Since thick disk stars observed in the
solar vicinity were born over a wide range of galactocentric distances (see section \ref{sec:data}),
their chemical properties should therefore reflect the variety in the
radial star formation histories of the thick disk, if any.

In the following, we show that these chemical properties are in fact
remarkably homogeneous at all ages in the thick disk, leaving little
room for some significant variation in the star formation history of
this population as a function of radius.  The outline of the paper is
as follows: in the following section, we briefly describe our data, in
Section \ref{sec:results}, we describe the model and present our results,
and discuss them in Section \ref{sec:discussion}.

\begin{figure}
\includegraphics[trim=40 60 0 80,clip,width=9.cm]{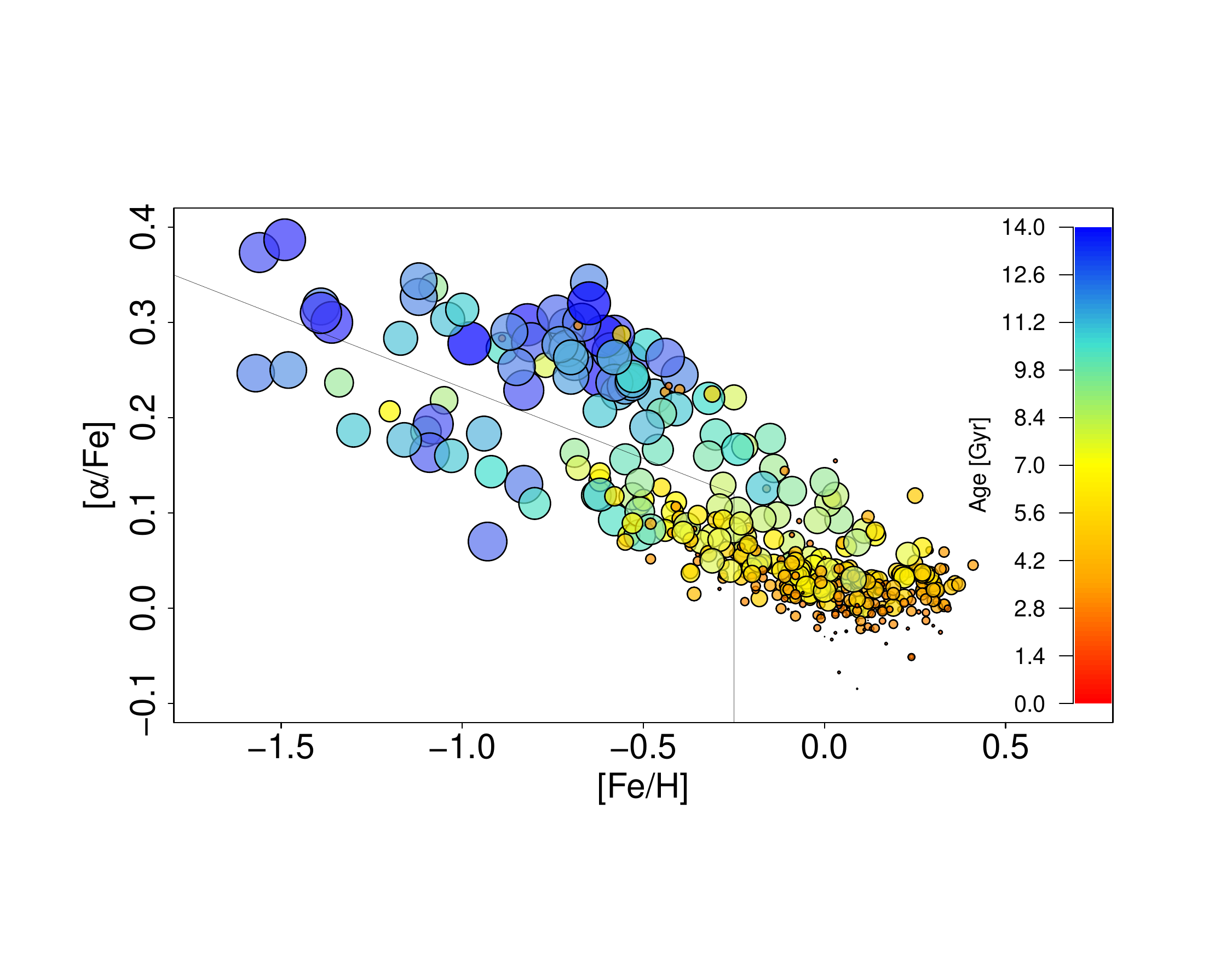}
\caption{The [$\alpha$/Fe]-[Fe/H] distribution at the solar vicinity for our sample. 
The colors (indicated in the legend) and size (larger implies older) of the symbols indicate the age of the stars.
The stars below and to the left of the line are either accreted stars according to \cite{nis10} 
(for those objects which have [Fe/H]$<$-0.7 dex), or metal-poor thin disk stars. They are represented by empty circles 
in Fig. 2, and removed from the age-alpha distribution in Fig. 3. 
}
\label{alphafehage}
\end{figure}

\begin{figure}
\includegraphics[trim=40 20 20 80,clip,width=9.cm]{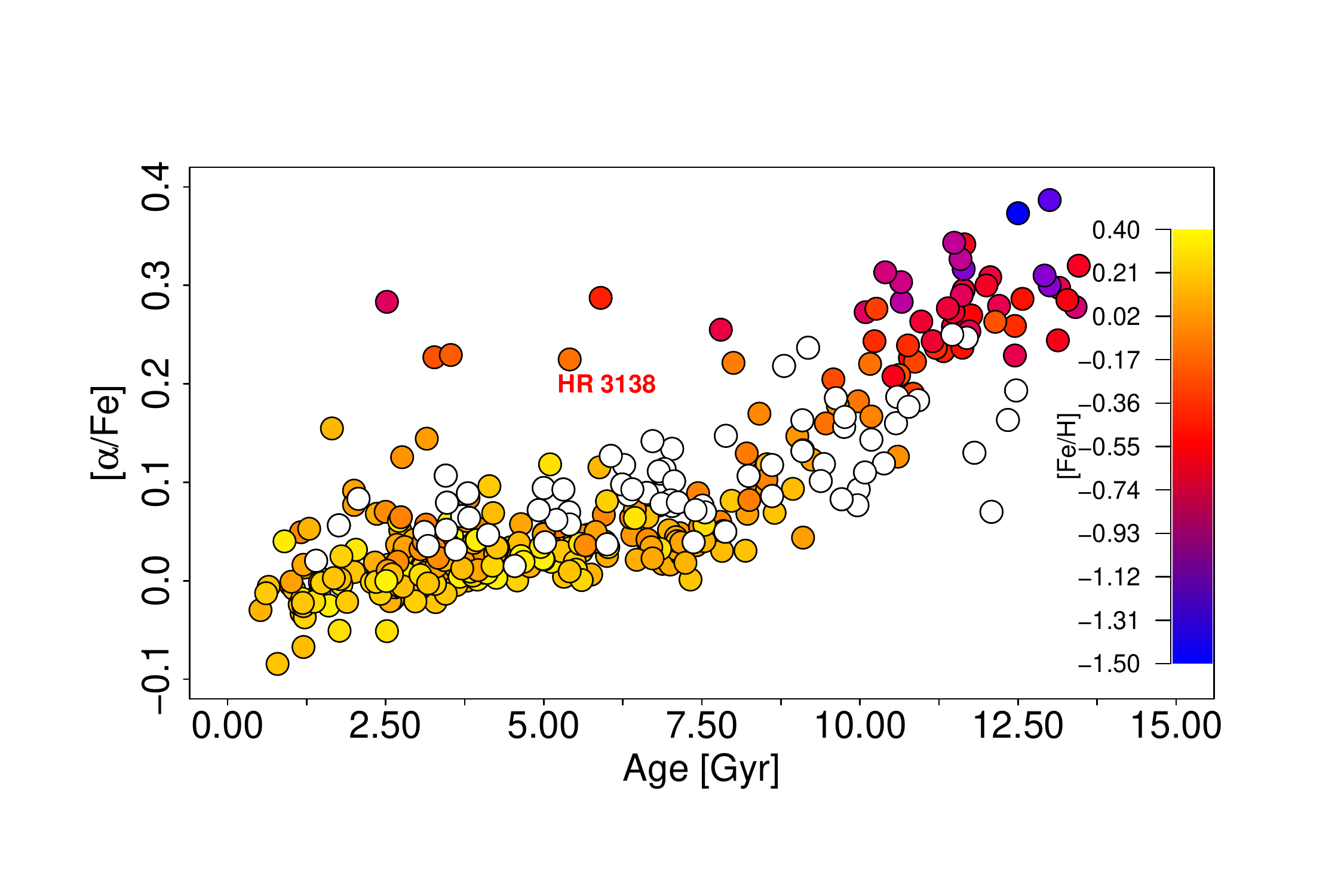}
\caption{The age-[$\alpha$/Fe] distribution of stars in Fig. 1.
Metal-poor thin disk stars and accreted stars, as selected from Fig. 1 (below and to the left of the limits in gray),
are shown as white dots.
}
\label{agealpha}
\end{figure}

\section{Data}\label{sec:data}

We use the sample of \cite{adi12} already analysed in \cite{hay13}.  
The sample contains 363 stars with age determinations, to which
we added further the 36 stars from the sample of \cite{nis10}
for which we could determine ages.  Ages for this last sample
were determined in the same manner as those from \cite{adi12}
sample in \cite{hay13}. However, as reported by \cite{sch12}, the effective 
temperature of the stars in Nissen \& Schuster may have been underestimated. 
We checked this statement by comparing the 11 Hipparcos stars in common between the two samples.
All these objects have a temperature lower in \cite{nis10} than in \cite{adi12}, 
with a mean offset of -75K, confirming the statement of \cite{sch12}.
Following their recommendation, we added +100K to the \cite{nis10} values.

In \cite{hay13} the spectroscopic scale of \cite{adi12} was compared to photometric temperatures
using the T$_{\rm eff}$--V-K calibration of \cite{cas10}, from which we deduced that the 
two scales are off by 56K. The impact of this offset on stellar ages of such difference in temperatures was studied in \cite{hay13}
and found to be limited (see their Fig. 2).
Ages were determined using the bayesian method of \cite{jor05}, and implies the derivation 
of a probability distribution function (or PDF) for each star. We adopted the Yonsei-Yale 
set of isochrones, with atmospheric parameters taken from the study of \cite{adi12},
including mean [$\alpha$/Fe] values. The uncertainties on the ages are derived from the PDF
as described in \cite{jor05}, and the values (see Fig. 3) are for a confidence
level of 68\%, corresponding to the $\pm$1 $\sigma$ Gaussian errors.
Random uncertainties on age reflect the uncertainties on the atmospheric parameters. 
In \cite{hay13}, we assumed 50K uncertainty on effective temperatures, noting 
that the formal uncertainties quoted by \cite{adi12} are sometimes much
smaller than this, 0.1 dex on metallicities, and 0.1 magnitude on absolute magnitudes.
The total uncertainties in the ages, as described in \cite{hay13}, are of-order 1 Gyr for the random error (with
a mean uncertainty of 1.1 Gyr), and a systematic error due to stellar physics of 1 to 1.5 Gyr.  The systematic uncertainties 
were estimated by comparing the ages determined using different sets of isochrones (see Haywood
et al. 2013 for details)
Given the dispersion of points on Fig. 3, of the order of 1 Gyr, we deemed our error estimates on age as realistic.

The [$\alpha$/Fe]-[Fe/H] distribution of the whole sample is shown on
Fig. 1.  

The stars below the separating line  
are considered either as metal-poor thin disk objects (see discussion
for these objects in \cite{hay08} and \cite{hay13} for the
status of these objects and \cite{sna14a,sna14b} for a description of
their chemical evolution), or as accreted stars, following the study by \cite{nis10}. 

\begin{figure}
\includegraphics[trim=40 60 0 80,clip,width=9.cm]{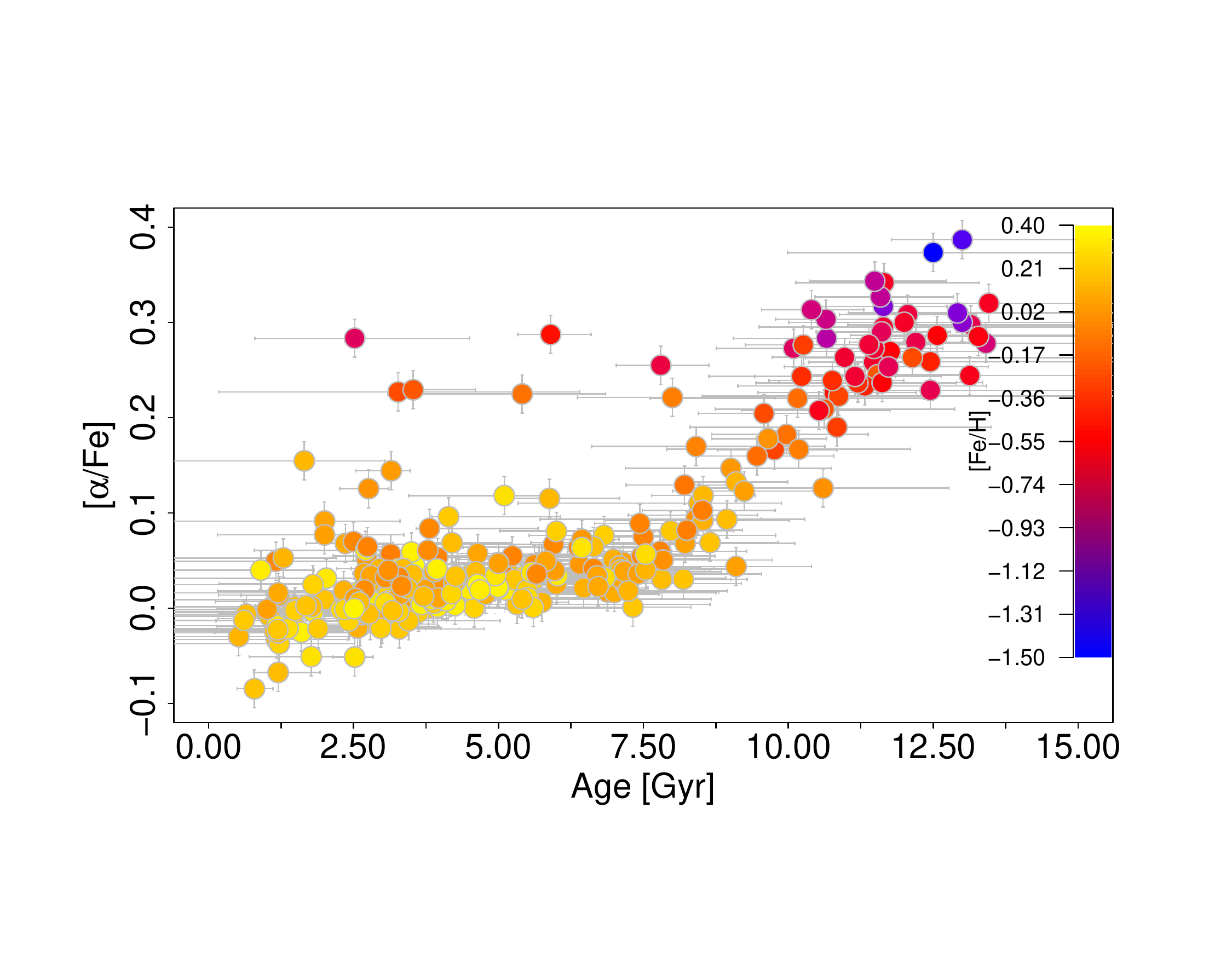}
\caption{The age-[$\alpha$/Fe] distribution at the solar vicinity after having removed metal-poor thin disk 
stars and accreted stars, as selected from Fig. 1. Note the remarkable small dispersion in both the thin and
thick disk sequences of stellar age and [$\alpha$/Fe]. Errors on age determinations are described in the text. 
On [$\alpha$/Fe], 0.02 dex error is taken as representative of the internal errors given in \cite{adi12}.
}
\label{fig:finalaa}
\end{figure}

In Fig. 2, which shows the age--[$\alpha$/Fe] distribution (the color
now coding the metallicity of each star), these objects are 
represented by empty circles.
The final distribution of Age-[$\alpha$/Fe],
after removing accreted and metal-poor thin disk stars, is plotted in
Fig. \ref{fig:finalaa}.  The sequences of the thin and the thick disks
appear clearly on the plot, with a sharp change of slope between the
two at $\sim$8 Gyr, as first discussed in \cite{hay13}.

Now more clearly demonstrated than in  \cite{hay13}, 
a number of objects with $\alpha$-enhanced abundance ratios have
age lower than the sequence of thick disk objects (age$<$8-9 Gyr at
[$\alpha$/Fe]$\sim$0.2-0.3 dex).  As we already discussed in  \cite{hay13}, these objects have no peculiarities.  We did not note at that
time however the dedicated study of \cite{fuh12} on HIP 38908
(HR 3138). \cite{fuh12} found similar ``young'' age to what
we derive for this object: we derive an age of 5.4$^{+1.4}_{-2.3}$ Gyr,
while \cite{fuh12} give 5.6$^{+2.2}_{-1.8}$ Gyr. They commented
further on the fact that radial velocity excludes the possibility that
HR 3138 is a blue straggler with a white dwarf.  The authors could
not find any specific reason for the young age, but suggested that
either the star had accreted gas from interstellar material or that the
enhanced effective temperature and luminosity resulted from a merger
with a companion\footnote{The other stars in our sample with similar
characteristics -- HIP 69780, 77637, 51191, 54469, 77610, 109822-- have
no known peculiarity.  Most (HIP 69780, HIP 77637, HIP 51191, HIP 54469
and HIP 109822) have clear asymmetric drift and overall hot kinematics,
confirming that they should belong to the thick disk or halo, in spite
of their age.  These objects are not considered further in the present
study. }.

In Fig. 4, we plot [$\alpha$/Fe] as a function of the pericenter
distance of stars.  The purpose of this plot is to illustrate that the
pericenters of thick disk stars span the whole range of radii from 8
kpc to the innermost regions. Hence, unless the orbital characteristics
of these stars were seriously affected by dynamical evolution, it can
be assumed that they represent a reasonable sampling of the thick disk
over a wide, perhaps even the whole, range of radial distances from the
Galactic center.

\begin{figure}
\includegraphics[trim=40 20 20 80,clip,width=9.cm]{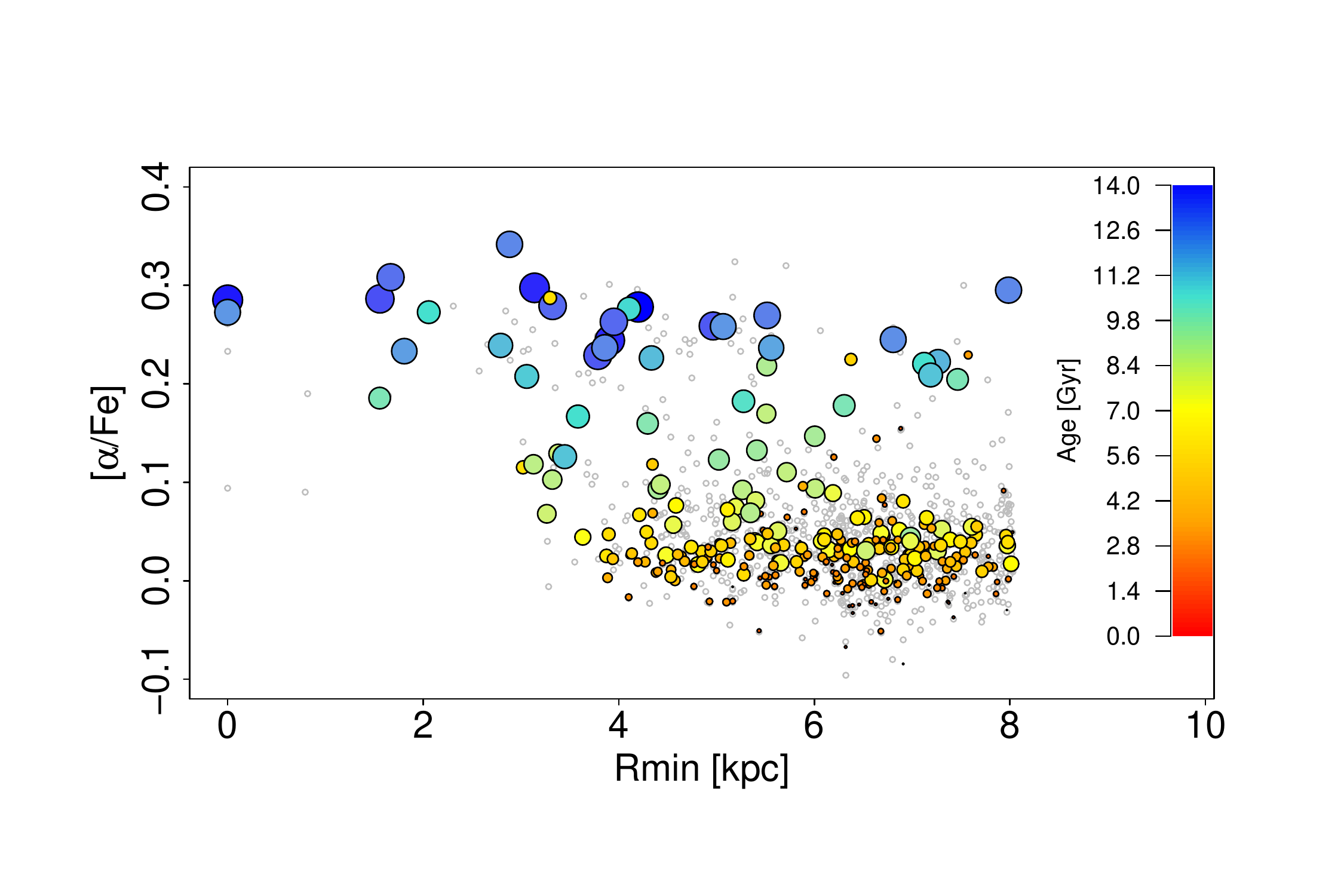}
\caption{Mean alpha abundance as a function of pericenter distance. Color (indicated in the legend) and size of the symbols 
indicate the age (larger symbols implies older ages) of the stars. Thick disk stars are essentially those stars with [$\alpha$/Fe]$>$0.1 dex.  
Grey points are for the whole sample. 
}
\label{fig:eccalpha}
\end{figure}

\section{Results: Modelling the [$\alpha$/Fe]-age relation}\label{sec:results}

\subsection{The model and chemical tracks}

The model is a closed-box and is discussed at length in \cite{sna14a,sna14b}.
Note that the effects described here are expected to be of similar, or even of higher
amplitudes with standard infall models. The reason being that since the ISM is 
less important in these models, they are more sensitive to the level of metal 
injection in the ISM than closed-box models, which have greater inertia in this
respect. A clear illustration of the effect of the variation of the SFHs on increasing 
the dispersion of alpha abundances at a given age is given in \cite{min13}, their
Fig. 7.

In the following, we use the
nomenclature of the disks as given in  \cite{hay13}.  That is,
we consider the thick and thin disk up to radial distances of 10 kpc
to form the inner disk, while stars beyond this limit are members of
the outer disk. We sketched a scenario for the formation of these two
different structures in  \cite{hay13} and the chemical evolutions
of both are described in \cite{sna14b}.

In short, the closed-box model assumes that:
(1) all the gas is in the system at the beginning. 
We take the closed-box model as the approximation of the case where most 
accretion in the inner disc (R$<$10 kpc) occurred early.
(2) The ISM is initially devoid of metals and remains
well mixed during its whole evolution.
(3) No coupling between the star formation rate and the gas
content is assumed. In \cite{sna14a,sna14b}, we determined the SFH by
fitting our model to the Age-[Si/Fe] data. Here we assume different star 
formation histories from which we derive the Age-[Si/Fe] tracks which
are compared to the data.

The ingredients of the model are: standard theoretical yields \citep{iwa99,nom06,kar10}, 
an IMF \citep{kro01}
independent of time, and a stellar mass--lifetime relation dependent
on the metallicity \citep{rai96}. No instantaneous recycling
approximation has been implemented, see \cite{sna14b} for details.

\subsection{Comparison with the data}

Models with radially dependent accretion time scales have a range
of star formation histories that vary from having a peak at early
times in the inner disk, to constant, or even increasing SFH in the
solar vicinity or beyond \citep[e.g.][]{naa06, min13,kub14}.  
Hence, in Minchev et al. the SFR at R=4kpc peaks at $\sim$
30 M$_{\odot}$.pc$^{-2}$.Gyr$^{-1}$ in the first Gyr, decreasing to
less than 10 M$_{\odot}$.pc$^{-2}$.Gyr$^{-1}$ at the present times,
while it is only slightly decreasing (by a factor of $\sim$ 2) at the
solar radius, and very nearly constant at 12kpc.  In Naab \& Ostriker,
the SFH at 4 kpc peaks a later time (6 Gyr), reaching similar values
(20-30 M$_{\odot}$.pc$^{-2}$.Gyr$^{-1}$), and decreasing to about 10
M$_{\odot}$.pc$^{-2}$.Gyr$^{-1}$ at the present epoch. The peak of the
SFR at the solar vicinity is reached only at $\sim$11 Gyr and thus is
mostly increasing throughout the evolution of the MW.  Independently of
any particular SFH, these models show that the SFR at early times varies
by large amounts depending on the radius: there is a factor of about 5
to 10 at 2-4 Gyr between the SFR in the inner regions (2-4 kpc) and the
outer regions (at about 8-10 kpc from the Galactic center).

\begin{figure}
\includegraphics[width=9.cm]{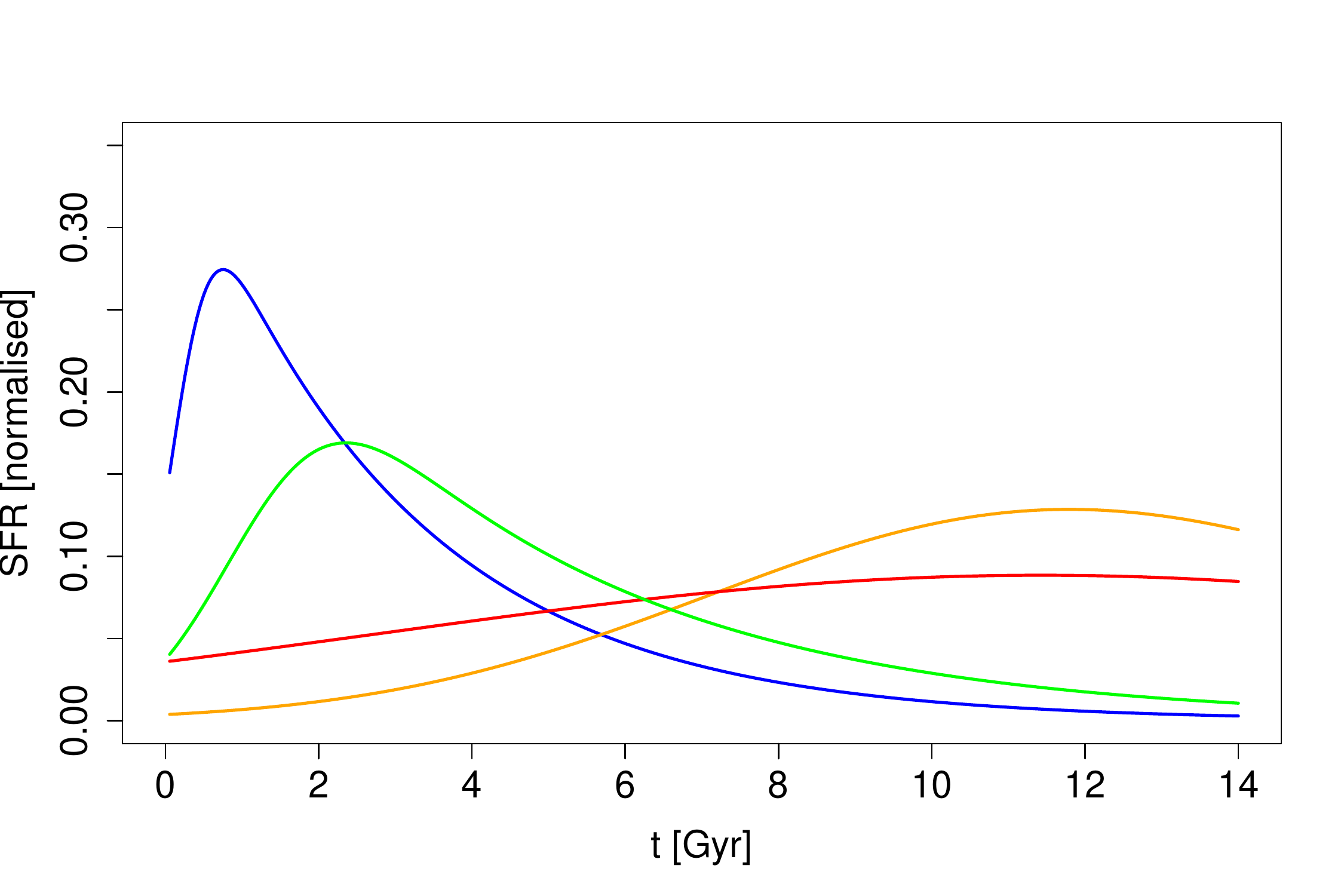}
\caption{The set of star formation histories used in this analysis.}
\label{fig:sfh}
\end{figure}

Because the radial variation of the SFHs can vary widely between one
model to the other, we do not rely on any particular model.  We have
chosen a set of SFHs which we take as representative of the general
phenomenology of radially dependent SFHs in disks, and that are shown
in Fig. \ref{fig:sfh}.  They are taken as illustrating the range of
variation that one can expect from adopting an infall time-scale and
related SFH varying with galactic radius: increasing SFH, constant SFH,
decreasing SFH, and reflect the 5-10 amplitude of the SFR in the first
few Gyr as noted above.

\begin{figure}
\includegraphics[width=9.cm]{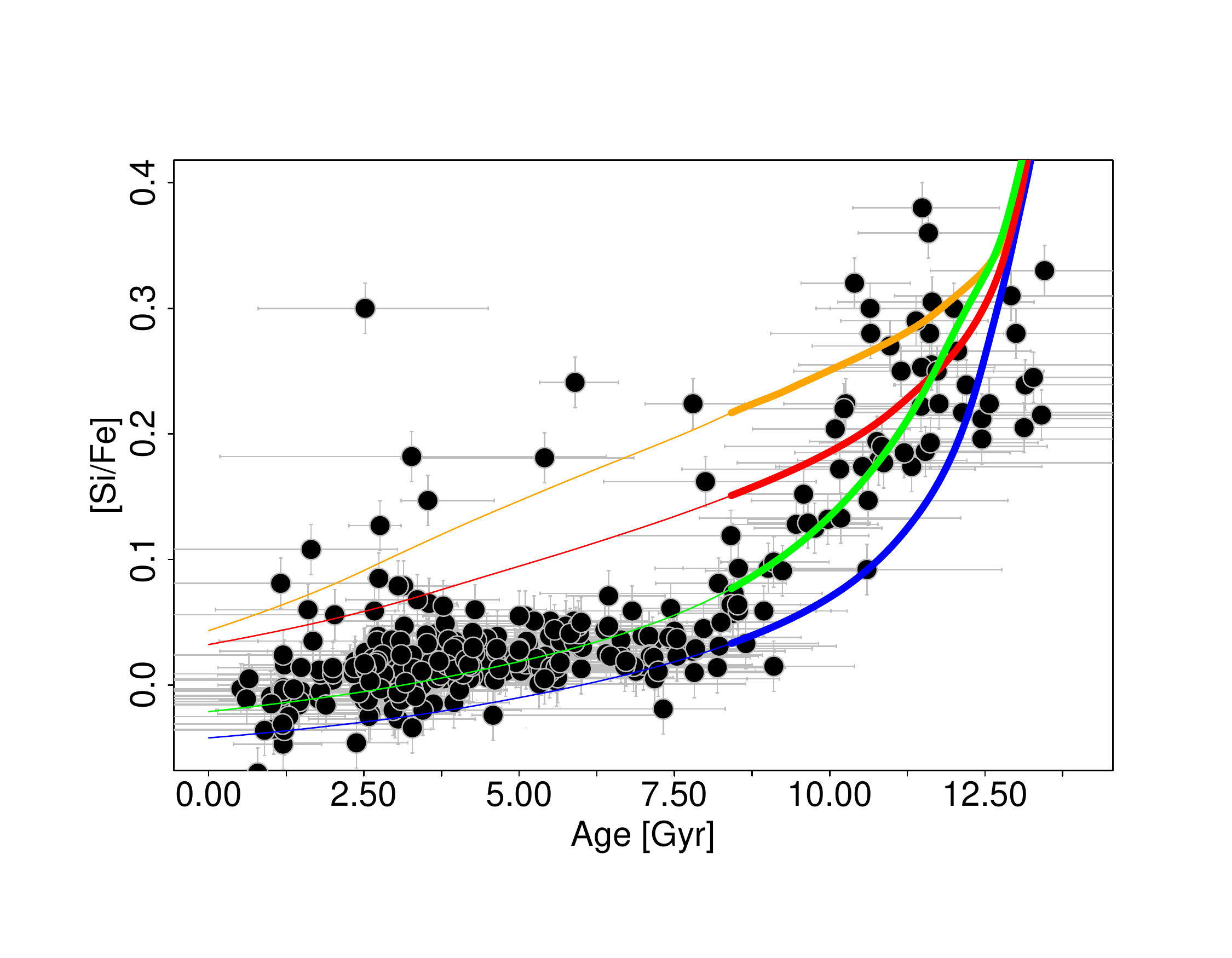}
\caption{The age $vs$ [Si/Fe] chemical evolution tracks corresponding to the SFHs given in Fig. 5, compared to the 
stellar age $vs$ [Si/Fe] data. The colors of each line correspond to those in Fig.~5. Thick part of the curves 
correspond to the thick disk phase. 
}
\label{fig:agesifeh}
\end{figure}

Fig. \ref{fig:agesifeh} shows the corresponding chemical tracks generated
from these SFHs over-plotted on the data.  As can be seen, the range
of alpha abundances spanned by the models at any given age is large,
bracketing the observed [Si/Fe] values.  As expected, the SFH that fits
best the observed age-[Si/Fe] distribution is one with a significant
peak of star formation at early times, similar to the one deduced by
fitting the data in \cite{sna14a}. If the peak in the SFH occurs
too early ($<$ 2 Gyr), the model generates too much iron early on and
the chemical track is too low compared to the data.  However,
if the SFH is flat (constant SFR), the knee at $\sim$ 8 Gyr is not
reproduced, with the chemical track being off the data by about 0.1 dex
over this region.  In the case where the SFR starts low and increases to
reach a maximum in the last few Gyrs as suggested by the model of 
\cite{naa06}for the solar vicinity, the model is far from the data.

As shown in the previous section, the pericenter distances of thick disk
stars span a large radial range from less than 2 kpc from the Galactic
center up to the solar circle (Fig. \ref{fig:eccalpha}).  Hence, we do
expect thick disk stars to sample the whole thick disk (R$<$ 10 kpc)
and their alpha abundance ratio to reflect  star formation histories
at a variety of radii.  Obviously, the expected dispersion is not
observed. The thick and thin disk sequences are very well defined and
at all ages, the dispersion in abundance ratio is small in both phases.
In fact, the observed dispersion at a given [Si/Fe] is fully compatible with 
the uncertainty of about 1~Gyr on the derived ages  \citep{hay13}, suggesting
the intrinsic dispersion in abundance is essentially negligible.  
In particular, we would expect the transition from the thick to the
thin disk at $\sim$8 Gyr to be much less well defined than it is. Had
the thick disk SFHs been different in the inner and outer regions, the
alpha abundance ratios at the end of the thick disk phase would show
a level of dispersion in their abundance ratios which is not observed.
As commented in  \cite{hay13}, the data strongly suggest that
the thick disk phase ended by leaving a disk of uniform properties from
which the thin disk started to form.

While the previous discussion was meant to be illustrative and useful for
comparing with SFH commonly used in models, the most pertinent question
is really: How much variation in the SFH is really allowed by the data?
This can be estimated by using the bootstrap SFHs generated in \cite{sna14b} 
to evaluate the uncertainty in the mean SFH that fits the data.
The bootstrap SFHs give us some kind of ``cosmic variation'' that is
allowed by the constraints provided by the data. What we seek here is not
the uncertainty in the SFR at a given time, but how much the SFH can be
systematically below or above the best fit SFH over a given time interval.
In order to do so, we summarize the information given by each of the 
bootstrapped SFHs by measuring the ratio of the stellar mass formed during
the thick disk phase (13-8 Gyr) to the total stellar mass formed (13-0
Gyr), for each solution. The mean and dispersion of these percentages
are 56 and 2.5\%, with values that vary, at most, between 50 and 61\%.
Note that the
derived uncertainty in the SFR at a given age is much larger than 2.5\% (see the error bars
in the derived SFH in \cite{sna14a,sna14b}),
but the overall variation of the SFR, as measured by the percentage of mass
created in the age range 13-8 Gyr relative to the total stellar mass, is much better constrained.
Comparatively, the stellar mass produced between ages 13 and 8 Gyr -- during the thick disk phase -- relative to the total 
stellar mass in each of
the model SFHs (Fig. 5) is 11, 26, 33 and 63\%. Hence, any significant variations 
from the mean SFH derived in \cite{sna14a,sna14b} are ruled out.

\section{Discussion}\label{sec:discussion} 

Summarizing the previous section, the observed dispersion in the age-[Si/Fe] distribution essentially 
reflects the uncertainties in the derived ages, although the thick disk stars originate from all over the disk. 
This implies that the chemical evolution in the thick disk phase 
must have been an extremely homogeneous process, incompatible with significant spatial variations of the 
SFR at these epochs. 
These is in clear contradiction with a number of models, which explicitly predict 
a very significant dispersion, see for example 
\cite{min13}, or \cite{kub14}, which both have a range of [$\alpha$/Fe] 
larger than 0.2 dex at the end of the thick disk phase.
Note however that this is likely to be true for all 
models which, motivated by the inside-out scenario, build there disk through a radially dependent SFH.

\subsection{The radially uniform SFH of the thick disk}

There are a number of important consequences
of this result.  In chemical evolution models based on long-time scale
accretion of gas \citep[e.g.][]{chi97, min13, kub14}, the SFH depends on the
infall history through some form of relation between the star-formation
intensity and the gas-surface density (i.e. a Schmidt-Kennicutt law). The
infall law is chosen in order to reproduce the phenomenology of inside-out
 galaxy formation models, and one of their possible signatures, chemical
 radial gradients.
Our results contradict a fundamental aspect of these scenarios: since the
SFH is measured to be the same across the (thick) disk, it may likely
imply that there is no radial dependence of the accretion of gas onto
the inner disk (R$<$10 kpc).  Although such simplistic inferences are
similar to those assuming direct connection between infall and star
formation intensity, they should be interpreted with caution however,
because as shown recently \citep{hop14,woo14}, 
the SFH does not necessarily correlate with the accretion history. Feedback
is expected to play an important role by reducing star formation rate
and efficiency thereby producing a reservoir of gas which can affect the
long-term fueling of the SFH of a galaxy.  Hence, infall could be radially
dependent, but not directly correlated to the SFR, implying that it cannot
be a determining parameter of the chemical evolution in the inner disk.

Another, perhaps more natural explanation, could explain the seemingly
uniform SFH in the thick disk.  In \cite{hay14a,hay14b} and \cite{sna14a,sna14b}, we provided a number of arguments to propose a new scenario
for the chemical evolution of the Milky Way disk. We argued that the
structural parameters and the measurement of the SFH of the Milky Way
provide new evidence that the thick disk is massive, and that this is
not expected in standard chemical scenarios (see the Introduction). The
only possible way to generate sufficient numbers of stars at intermediate
metallicities (which, we argued, are mostly confined to the inner regions,
as one would expect for a population with 2 kpc scale length, \cite{ben11}, \cite{che12a}, \cite{bov12}), is that
the gaseous disk was already massive at early times.  From the point
of view of chemical evolution, the model that is the most likely to
represent this situation is a closed-box model.  In such case, because
of the short scale length of the disk, the relative large scale of the
turbulence \citep{leh14}, and the absence of radially dependent
accretion time scale, less differentiation in the SFHs is expected to
take place in the inner and outer regions of the thick disk.

\cite{nid14} confirm this conclusion on the basis of the
analysis of the APOGEE data and a chemical evolution model.  Their model
includes infall with a long time scale, but no radial dependence, hence,
no induced variation in the SFH due to some kind of inside-out formation.
They assessed however possible variations in the SFH by assuming
different star formation efficiency, and deduced that some homogeneity
in the SFH can be deduced from the limited dispersion observed in the
[$\alpha$/Fe]-[Fe/H] distribution in the thick disk regime sampled by the
APOGEE data.  As in  \cite{hay13} and \cite{leh14}, they
note that this implies that the thick disk has formed from a turbulent,
well-mixed gaseous disk.

\subsection{On the inside-out galaxy formation model}

Other evidence supports a uniform SFH for the thick disk.
First, no radial metallicity gradients have been found in the thick
disk \citep{che12b}. This is expected if the SFH is radially
undifferentiated.  Note that the tight correlation found between age
and chemical parameters (see  \cite{hay13}, and previous section)
in the thick disk implies that the lack of gradient is not a consequence
of mixing in the {\it stellar} disk, but of a well mixed ISM at a given
epoch in the thick disk  \citep{hay13}. Chemical properties of
the thick disk evolved homogeneously over time.  We already commented
in  \cite{hay13} and  \cite{leh14} that this was most
probably a consequence of the high turbulence in the gaseous disk at
these epochs. Turbulence induces sufficient mixing to explain the high
homogeneity of chemical species in the ISM \citep{fen14}.
Stars formed out of this well-mixed gas in a geometrically thick layer.

Second, no dependence of the radial scale length of the thick disk as
a function of the $\alpha$-abundance ratios has been detected (see our comments
in  \cite{hay13} on the Fig. 5 of \cite{bov12}).  \cite{hay13} demonstrated that alpha abundances
correlate well with age in the thick disk phase (see also the previous
section).  Hence, this implies that no significant dependence of the
scale length with age, or an inside-out age trend during the formation of
the thick disk, is detected.  This is in accordance with the fact that
no significant variation of the SFH is imprinted in the age-$\alpha$
relation, which we would expect if the youngest populations had the
longest scale lengths.  This is also in good accord with the general
scheme presented in  \cite{hay13} where we discussed the
inside-out galaxy formation scenario.  

Note that the Haywood et al. study has been cited as claiming evidence
that the MW did not form inside-out. We caution that this is an
oversimplification of the view presented in that paper.  What we claimed
is that there is no such structure as a single disk formed smoothly and
continuously from the inside-out and generating smooth radial chemical
gradients.  Indeed, there are two disks in our nomenclature (see  \cite{hay13}): the inner one (R$<$9-10kpc), composed of the thick and
the thin, with different scale lengths\footnote{Note that the value of 3.6~kpc is a representative mean, but at
[Fe/H]$>$-0.1 dex the scale length measured by \cite{bov12} for the thin disk is a complex function 
of metallicity and alpha abundance. For instance, metal-rich stars, which are more likely to belong 
to the inner thin disk have a scale length significantly shorter than the value of 3.6~kpc.
This complex issue is discussed in Haywood et al. (2015, in preparation).}($\sim$2 and 3.6 kpc, see \cite{bov12}), but with no evidence of an increase in the scale length as
each of these disks grew.  And there is the outer disk, which started to
form before the inner thin disk, 10 Gyr ago, and which has a long scale
length ($\sim$ 4 kpc or more).  Hence, the general trend at
early times ($>$ 8 Gyr) would be mostly imprinted by the formation of the massive,
and concentrated, thick disk, followed by the outer (but less dense) disk
which has a much longer scale length.
Hence, snapshots of such a galaxy at different epochs
would certainly give the impression of an inside-out growth of its disk,
although the overall sequence is more complex and is not a smooth process.
What we did claim in Haywood et al. and now argue further is that this
overall scenario is in contradiction with the simplest picture
of inside-out galaxy formation as we now discuss.

\subsection{The formation of disks}

Within the context of these results, models of formation of MW-like
galaxies may need some, perhaps significant, revisions.  Models where
the radial dependent gas accretion drives a sequence of star formation
which is inside-out during the formation of the thick disk are now likely
ruled out.  In addition, models, at least during the phase of thick disk
growth, need to be such that the exponential disk scale length does not
increase. \cite{fer01} explored how the scale length of
disks is affected during infall when the time scales for star formation
and viscous gas redistribution are similar (within a factor of a few).
They show that, if the main infall phase precedes the onset of star
formation (the equivalent of a closed box model), the scale length is
invariant with time, because the scale length is determined by the mean
specific angular momentum of the gas already settled and viscosity
redistributes the angular momentum such that the scale length is
invariant even as the total extent of the disk grows.  On the contrary,
if the infall time scale is similar to the lifetime of the disk, the
scale length evolves according to the specific angular momentum of the
accreted gas.  Hence, at first order, a non-evolving scale length in
the thick disk is a natural outcome of a pre-existing disk rich in gas.
Interestingly, Ferguson \& Clarke predict that, in their pre-assembled
disk, ``half of the stellar mass is in place more than 6 Gyr ago'',
in good agreement with the SFH measured by \cite{sna14a,sna14b}.

The advantage of viscous dissipation is that it naturally forms
exponential disks with a large span in radius \citep{fer01}
and because the viscous time scale is similar to that of the
star formation timescale\footnote{The viscosity is given by $\nu$ =
v$_{\rm turb}$ l$_{\rm turb}$, where v$_{\rm turb}$ is the turbulent
velocity of the largest energy injection scale, and l$_{\rm turb}$
is this largest scale. The viscous time scale is simply, t$_{\nu}$ =
R$^2$/$\nu$ \citep{vol11}. Scaling this relation to observations
of z=2-3 galaxies and values consistent with the thick disk of the MW
\citep{bov12,hay13,leh14}, suggests,
$\nu$ = 0.05 (v$_{\rm turb}$/50 km s$^{-1}$) (l$_{\rm turb}$/1000 pc)
pc$^2$ yr$^{-1}$ and t$_{\nu}$ = 6$\times$10$^7$ (R/1.8 kpc)$^2$ yr.
The global rate at which stars form is approximately one dynamical
time of the thick disk, t$_{\rm dyn}$= 6$\times$10$^7$ (R/1.8 kpc)
(v$_{\rm rot}/200 km s ^{-1}$)$^2$ yr.  These crude estimates suggest
that plausibly, t$_{\nu}$$\approx$t$_{\rm dyn}$ or similar to the star
formation time scale of the disk, and thus viscosity may be important for
redistributing the gas.}, it helps to alleviate the problem of forming
massive bulges which plague simulations without strong stellar feedback \citep{dut09}.
Viscosity both moves gas outwards,
extending the disk, and prevents gas from accumulating into the center
of the potential.  
Since the disk scale length does not increase when both viscosity is important, 
and the accretion and star formation timescales are similar, gas is distributed
with the same radial dependence as the previous generation of stars.

Thus viscosity is able to prevent the thick disk from developing a radial metallicity gradient or increasing 
its scale length with time or with increasing metallicity. 

 We note that the analysis of \cite{vol11} suggests that star formation in local disk galaxies with stellar mass M$_{\star}$ $\la$10$^{10}$ M$_{\sun}$
may also be sustained by the viscous dissipation of the (outer) HI extended disks.
Viscous dissipation allows the gas to move inwards and distributes it radially within the disk at rates
sufficient to support the observed rate and distribution of star formation.
Their analysis of local disk galaxies and our results imply that viscous dissipation
and gas redistribution may be generically important to the growth of disks.

Gas accretion, regulated by the intensity of on-going star
formation, may help to alleviate both the problem of needing the gas
accretion to follow the radial distribution of the star formation and of maintaining a constant
disk scale length.  For example, \cite{mar11} and \cite{mar12}
have suggested gas driven into the hot coronae by recent 
star formation in the disk may cause the halo gas to cool.
This cool gas can then accrete onto the galaxy.
Since the rate at which mass and momentum are injected in the halo gas controls 
how much gas is ultimately accreted, and these injection rates depend on the star formation intensity of the disk, 
generates a self-regulating cycle whereby more intense star formation leads to more intense accretion,
resulting in further star formation, and so on. 

In such a model, unlike the recent models of gas accretion from the cosmic web 
\cite[e.g.][]{dan14}, the gas accretion rate would follow the radial intensity of
on-going disk star formation and, perhaps in conjunction with viscosity,
maintain a constant disk scale length.

The results of the study by \cite{van13}fits well in this
scheme, first because of a similar mass growth found on progenitors
of the Milky Way (see \cite{sna14a}, \cite{leh14}), but
also by showing that their galaxies do not increase their size through
inside-out growth between z=2.5 and 1, but through self-similar mass
growth at all radii (see also \cite{mor15}).  Similarly, \cite{nel12} find that galaxies
begin to form inside-out after z=1 (see comment by \cite{van13}). This is precisely what is predicted
by viscous redistribution of the gas \citep{fer01}, and in
agreement with our findings here.

\section{Conclusions}

We have studied the impact of varying the SFH on the age-[$\alpha$/Fe]
relation of disk stars.  We show that a radially dependent growth of
the disk would give rise, through a radial variation of the SFHs, to a
dispersion in this relation which is not observed. 
Star formation in the thick disk was a general, large scale process, that 
led to a homogeneous chemical evolution throughout the inner disk (R$<$10kpc).
Consequences of these results for chemical evolution models have been explored. 
If infall intervenes
on a long time scale, as advocated in standard chemical evolution model, it
implies either that infall is not varying with Galactocentric distance
in the inner disk, or that infall is decoupled from the SFH. In any
case, it implies that infall is not a determining parameter of the
chemical evolution of the inner disk.  In \cite{hay14a,hay14b}, \cite{sna14a,sna14b}, we have shown that the characteristics of the thick disk
favor a scenario where stars started to form from a disk rich in gas
(most accretion occurred before significant star formation took place),
which we model as a closed-box.  We argued that the present results
find better justification within this new context.  It is reasonable to
expect that a uniform SFH is more likely to occur in a disk with a short
scale length and given the high level of turbulence in disks at high
redshift, and with no radially dependent infall time scale. Moreover,
differentiated SFHs are expected to produce both radial chemical
gradients and an increase in the scale lengths of the disk, expected in
inside-out formation scenarios, while none are seen in the observations.
The study by \cite{fer01} is qualitatively in accord with
our results: they argue that, under the assumption of similar time scale
for star formation and viscosity, in a pre-assembled disk, no significant
variation of the scale length is expected, and therefore no inside-out
growth of the disk.

\begin{acknowledgements}
The authors gratefully acknowledge the referee for helpful and constructive comments on the paper.
This research has made use of the SIMBAD database, operated at CDS, Strasbourg, France 
\end{acknowledgements}

\clearpage

\end{document}